\documentclass[aps,prl,twocolumn,amssymb,superscriptaddress]{revtex4}
\usepackage{subfigure}
\usepackage{graphicx}
\usepackage{rotating} 
\usepackage{color}
\usepackage{float}
 \usepackage{amsmath,amsthm}
\usepackage{epstopdf}
\begin{document}
\title{Time traces of individual kinesin motors suggest functional heterogeneity}
\author{Babu J. N. Reddy}
\affiliation{Department of Developmental and Cell Biology, University of California Irvine, Irvine, CA 92697, USA}
\author{Suvranta Tripathy}
\affiliation{Department of Developmental and Cell Biology, University of California Irvine, Irvine, CA 92697, USA}
\author{Jing Xu}
\affiliation{School of Natural Sciences, University of California, Merced, California 95343, USA}
\author{Michelle Mattson}
\affiliation{Department of Developmental and Cell Biology, University of California Irvine, Irvine, CA 92697, USA}
\author{Karim Arabi}
\affiliation{Department of Developmental and Cell Biology, University of California Irvine, Irvine, CA 92697, USA}
\author{Michael Vershinin}
\affiliation{Department of Physics and Astronomy, University of Utah, Salt Lake City, UT 84112, USA}
\author{Steven Gross}
\thanks{sgross@uci.edu}
\affiliation{Department of Developmental and Cell Biology, University of California Irvine, Irvine, CA 92697, USA}
\author{Changbong Hyeon}
\thanks{hyeoncb@kias.re.kr}
\affiliation{Korea Institute for Advanced Study, Seoul 130-722, Korea
}
\date{\today}

\begin{abstract}
Conventional analysis of in vitro assays of motor proteins rests on the assumption that all proteins with the same chemical composition function identically; however molecule-to-molecule variation is often seen even in well-controlled experiments. 
In an effort to obtain a statistically meaningful set of time traces that simultaneously avoid any experimental artifacts, we performed quantum-dot labeled kinesin experiments on both surface and levitated microtubules. 
Similar to glassy systems, we found that mean velocities of individual kinesin motors vary widely from one motor to another, the variation of which is greater than that expected from the stochastic variation of stepping times. 
In the presence of heterogeneity, an ensemble-averaged quantity such as diffusion constant or randomness parameter
is ill-defined. 
We propose to analyze heterogeneous data from single molecule measurements by decomposing them into homogeneous subensembles. 
\end{abstract}

\maketitle

Single molecule experiments in the last decades have greatly increased our understanding of biomolecules through real-time visualization of molecular movement \cite{Visscher99Nature,ZhuangSCI02,Shubeita08Cell}, making it possible to perform time series analysis for individual molecules, as well as to calculate the distribution of dynamic variables that had previously been difficult to determine in ensemble-averaged measurements.   
One of the most striking, yet not well appreciated, observations from single molecule measurements is that there are persistent heterogeneities at the \emph{molecular} level in a number of biological systems \cite{Xie98Science,Hyeon2012NatureChem,Liu2013Nature,ZhuangSCI02,Solomatin10Nature}.
In a system with molecular heterogeneity, even if all the molecules are chemically identical and under the same experimental conditions, a dynamic pattern observed in one molecule differs substantially from other molecules and this variation is greater than the stochastic variation expected from the mean. 

According to the general principle implied by the thermodynamic hypothesis in molecular biology (Anfinsen's dogma) \cite{Anfinsen73Science},  (i) native states of at least small proteins, which can also be extended to RNA \cite{Thirum05Biochem}, are uniquely determined by a given sequence and external condition, and 
(ii) dynamics of biomolecules occurs reversibly with high kinetic accessibility. 
Both conditions are regarded to be critical for biomolecules to achieve functional fidelity. 
In the light of such principles, it is rather surprising to find that variation of dynamical property is greater than the variation allowed by the stochasticity of rate process, and such heterogeneity persists far longer than a biologically viable time scale \cite{Solomatin10Nature,Hyeon2012NatureChem}.    
Recent single molecule studies, such as the docking-undocking transition of ribozymes \cite{ZhuangSCI02,Solomatin10Nature} and  isomerization dynamics of Holliday junctions \cite{Hyeon2012NatureChem}, clearly show that dynamics of molecules in action can differ drastically from one another.  
This puzzle has recently been illuminated by the variable mean velocity of the DNA helicase, RecBCD \cite{Liu2013Nature}.  
Formally, kinetics of a single enzyme is stochastic, so it is a priori not obvious whether the variation simply reflects such stochasticity. 
Critically, the theoretical analysis to understand such heterogeneity is poorly developed.
Here, we discuss 
when the observed single-molecule variation is too large to be generated from molecules functioning with the same underlying kinetic parameters. We then apply it to kinesin, and discover that such heterogeneity exists in the single-molecule function of the kinesin motors.
 
\begin{figure*}[t]
\includegraphics[width=6.8in]{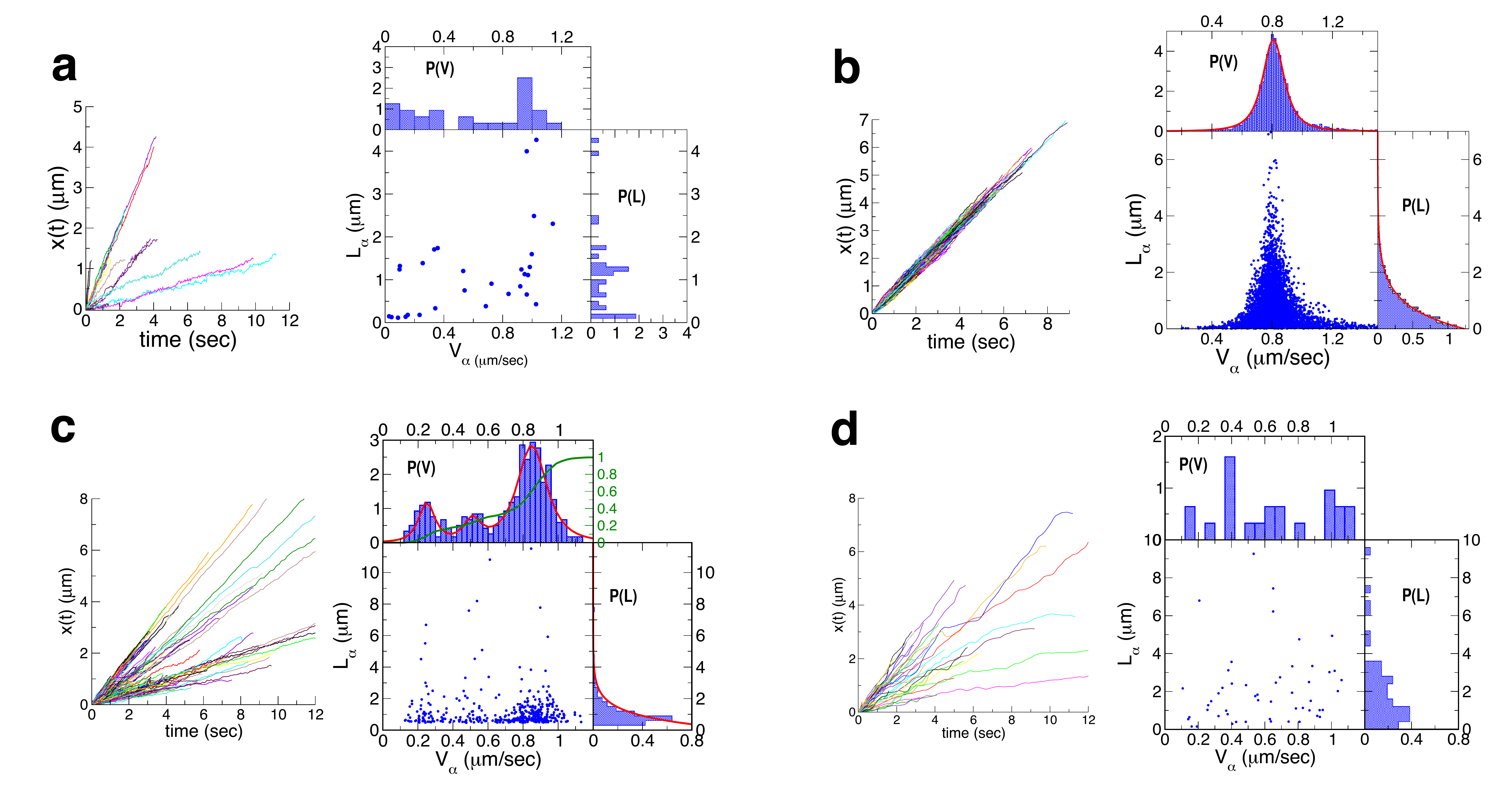}
\centering
\caption{
(a) Time traces of single kinesin motors from bead assays ($N=32$). 
The tail domain of individual motor is specifically attached to the bead through strepavidin-biotin with 30 \% of kinesin/bead binding fraction \cite{vershinin2007PNAS}. 
Scatter plot of ($V_{\alpha}$,$L_{\alpha}$) is shown on the right with the corresponding histograms. 
(b) Computer generated time traces ($N=5000$) with $\psi(t)=\tau^{-1}e^{-t/\tau}$ ($\tau=10$ ms) and detachment probability of $p=0.01$ at each step. The histogram of travel distance (right panel) is fitted to $P(L)=\overline{L}^{-1}e^{-L/\overline{L}}$ with $\overline{L}=0.80$ $\mu m$ (red line); and the velocity distribution (top panel) is to Eq.\ref{eqn:PV} with $\overline{D}=0.0033$ $\mu m^2/s$, $\overline{V}=0.81$ $\mu m/s$, and $\overline{L}=0.89$ $\mu$m (red line). 
(c) QD-labeled kinesin data on surface-immobilized MTs ($N=397$). 
The velocity distribution is fitted to Eq.\ref{eqn:three_pop} (red line).  
The green line depicts the cumulative sum of the histogram. 
Meanwhile, the histogram of travel distance (see Fig.S3 for unnormalized one) is fitted from the second bin because the trajectories with short run length, which contributes to the first bin, are excluded from the velocity analysis, satisfying $P(L)=\overline{L}^{-1}e^{-L/\overline{L}}$ with $\overline{L}=0.78$ $\mu m$.
(d) QD-labeled kinesin data on levitated MTs ($N=52$).  
\label{fig:velocity}}
\end{figure*}

To preclude the possibility that the observed heterogeneity is an undesired outcome of experimental artifacts, we performed a few distinct control experiments. 
(i) In bead assays, nonspecific interaction between bead and the kinesin tail domain could lead to the heterogeneous dynamics. 
To avoid this possibility we attached a bead to the kinesin tail domain via a specific streptavidin-biotin interaction (see SI).
The trajectories generated from the bead assays ($N$=32, Fig.\ref{fig:velocity}a) vividly 
show velocity variations consistent with motor heterogeneity. 
To determine if the heterogeneous ensemble of motors could be partitioned into functional sub-classes, we developed a new QD/TIRF experimental setup (Fig.S1), allowing improved ($N$=397) statistics. 
Importantly, in our TIRF-visualized quantum-dots (QDs), to which the motor's tail domain is specifically attached via an antibody, the motors (K560) lack the auto-inhibitory portion of the tail \cite{kaan2011Science}, so any observed variability does not reflect variable auto-inhibition. 
Again, velocity variations were observed, consistent with motor heterogeneity that is intrinsic to the individual molecules, and suggests a need to reconsider many facets of the conventional single molecule data analysis. 

Kinesin-1 is a motor protein that ``walks" processively along microtubules by converting chemical potential of ATP hydrolysis into mechanical stepping \cite{Visscher99Nature}. 
At saturating ATP condition, 
it is often stated that kinesins move at mean velocity of $\overline{V}\approx$ 0.8 $\mu m/sec\text{ (=8 nm/10 ms)}$
\cite{Visscher99Nature}. 
However, such a  description does not make explicit whether the average velocity of 0.8 $\mu m/sec$ is a common property of all the individual kinesin molecules or it is simply a property calculated over the heterogeneous population.

The actual time traces of kinesins (K560) from bead assays 
display substantial heterogeneities in the mean velocity (Fig.\ref{fig:velocity}a). 
This large dispersion seems unlikely to be explained by stochasticity of stepping time comprised of binding of ATP, hydrolysis, and ADP release.  
While kinesin motors remain on microtubule tracks, slow motors are persistently slow and fast motors are persistently fast. Slow-to-fast or fast-to-slow interconversion of velocity is not observed on the time scale that a kinesin remains on its track. 
Since each realization of a kinesin time trace is ideally a consequence of the cumulative sum of stochastic yet uncorrelated steps, one can assume that the stepping time is drawn from an ``independent and identically distributed"  (i.i.d.) random variable.  
In this case, due to the elementary renewal theorem \cite{vanKampen}, 
any kinesin molecule ($\alpha=1,2,\cdots N$) should have the \emph{identical} mean velocity at long time (or sufficiently large number of steps), 
$t\rightarrow\infty$, i.e., $\lim_{t\rightarrow\infty}V_{\alpha}(t)=\overline{V}$ for any $\alpha$.  
Evidently, even when the number of steps is large enough ($t/\tau\sim \mathcal{O}(10^2) \gg 1$), the time traces in Fig.\ref{fig:velocity}a show persistent and distinct motor movement leading to distinct mean velocities, thus violating LLN as in glassy systems \cite{Thirumalai89PRA,kang2013PRE}. The assumption of i.i.d. is not valid among the kinesin time traces.

To makes this point more straightforward, let us consider an ensemble of Poisson walkers that take steps with a stepping time distribution $\psi(t)=\tau^{-1}e^{-t/\tau}$.  
The positions of walkers along the track are expected to evolve as
$\partial_tx(t)=\overline{V}+\eta(t)$ where $\eta(t)$ is Gaussian white noise obeying $P[\eta(t)]\propto e^{-\int^t_0d\tau \eta^2(\tau)/4\overline{D}}$ with $\langle \eta(t)\eta(t')\rangle=2\overline{D}\delta(t-t')$; then the probability of finding a walker at position $x$ at time $t$ is  
$P[x(t)]=\left(4\pi \overline{D}t\right)^{-1/2}\exp{\left[-(x(t)-\overline{V}t)^2/4\overline{D}t\right]}$,
and hence the mean velocity $x(t)/t=V(t)$ up to time $t$ should obey the gaussian-like velocity distribution 
\begin{equation}
P_G[V(t)]=\left(\frac{t}{4\pi \overline{D}}\right)^{1/2}\exp{\left[-\frac{(V(t)-\overline{V})^2}{4\overline{D}/t}\right]}. 
\label{eqn:Gaussian}
\end{equation}
The variance of $V(t)$ decreases as $t^{-1}$, and the distribution finally converges to $\delta(V(t)-\overline{V})$, suggesting that after a sufficient number of steps, any walker ought to have an identical mean velocity $\overline{V}$. 
Indeed, an ensemble of simulated time traces with $\tau=10$ ms and $d=8$ nm confirms that the mean velocities from an ensemble of Poisson walkers at time $t$,  are distributed as a gaussian, centered around 
$\overline{V}=0.8$ $\mu$m/s (Fig.S2a). 
The same conclusion is reached even for a more general stepping time distribution, though a single rate limiting step ($\psi(t)\sim e^{-t/\tau}$) is a relatively ``worst-case" scenario, generating relatively large amounts of heterogeneity in velocity. 
As an example, a stepping time $\tau$ composed of multiple internal substeps ($\psi(t)\sim t^ne^{-t/\tau}$) leads to less dispersed time traces (Fig.S2b).  
As long as the stepping time is i.i.d. and has a \emph{finite variance} ($\langle(\delta t)^2\rangle<\infty$), the above conclusion, i.e., $P(V(t))=\delta[V(t)-\overline{V}]$ as $t\rightarrow \infty$ is always valid, which simply restates the central limit theorem.

In practice, kinesin motors have a finite processivity satisfying exponential travel time distribution, 
$P_{\overline{L}}(t)=(\overline{V}/\overline{L})e^{-\overline{V}t/\overline{L}}$, as is the case in our simulation result where a detachment probability ($p=0.01$) is imposed on each step (see Fig.\ref{fig:velocity}b). 
For an ensemble of \emph{homogeneous} motors with exponential travel distance, the distribution of mean velocities ought to be described by incorporating the travel time distribution as a weighting factor, which leads to:
\begin{align}
P_{\text{homo}}(V&;\overline{V},\overline{D},\overline{L})=\int^{\infty}_0dtP_G[V(t)]P_{\overline{L}}(t)\nonumber\\
&=\frac{\overline{V}/\overline{L}}{4\sqrt{\overline{D}}}\left[\frac{(V-\overline{V})^2}{4\overline{D}}+\frac{\overline{V}}{\overline{L}}\right]^{-3/2}.
\label{eqn:PV}
\end{align}
$P_{\text{homo}}(V;\overline{V},\overline{D},\overline{L})$ is symmetric with respect to $\overline{V}$ with a single peak and power-law tails at both ends. 
It is of particular note that from the scatter plot of $(V_{\alpha},L_{\alpha})$ (Fig.\ref{fig:velocity}b) traces with large run length is found predominantly near the mean velocity, and the traces showing the large deviation from $\overline{V}$ always have a short run length.  
The full width at half maximum (FWHM) of Eq.\ref{eqn:PV} is $2(2^{2/3}-1)^{1/2}(\overline{V}\overline{D}/\overline{L})^{1/2}\approx 3.1(\overline{V}\overline{D}/\overline{L})^{1/2}$. 
The predicted velocity distribution is indeed realized in velocities calculated from simulated Poisson walkers (see Fig.\ref{fig:velocity}b); however, we observe that it fails to adequately describe our experimental data in a number of ways. 

Although the violation of LLN is obvious among the time traces from the bead assays (Fig.\ref{fig:velocity}a), with only $N=32$ traces we could not investigate whether there might be distinct subgroups of motors with similar function.  
Thus we also used QDs (Fig.S1) to monitor the movement of individual kinesin motors; this experimental geometry not only gave more traces, but was perhaps more natural than the bead/optical trap assays that bring motors into contact with the MT because individual QD-labeled motors land spontaneously on MTs and start to move. 

In qualitative agreement with the bead assay data, the time traces and $P(V)$ with multiple peaks from TIRF-visualized QD data show the motor heterogeneity more convincingly (Fig.\ref{fig:velocity}c). 
The time traces in Fig. \ref{fig:velocity}c clearly contradict the usual assumption of i.i.d., suggesting that not all the kinesin motors are functioning identically, and that the energetic cost of overcoming such heterogeneity is not minor. 
One might wonder whether interaction between either the QD or motor nearby surface to which the MTs are attached can account for the observed heterogeneity. 
If the surface interaction were significant, the velocity of kinesin would vary depending which protofilament the kinesin walks along. 
To preclude this possibility, we performed QD-assays on MTs levitated above the surface. 
Evident from Fig. \ref{fig:velocity}d, we still find the motor heterogeneity even in the QD-labeled kinesin time traces from the off-surface measurements. 
The mean velocities of motors on the levitated MTs (Fig. \ref{fig:velocity}d) are widespread, demonstrating qualitative agreement with those on the surface-immobilized MTs.

Eq.\ref{eqn:PV}, that assumes homogeneity of motors, fails to explain the overall asymmetric and multiply peaked velocity distribution of our QD-labeled kinesin data in Fig. \ref{fig:velocity}c.
Thus, we propose a \emph{heterogeneous} velocity distribution by combining homogeneous subpopulations:
\begin{align}
P_{\text{hetero}}(V)=\sum_{i=1}^n\phi_iP_{\text{homo}}(V;\overline{V}_i,\overline{D}_i,\overline{L})
\label{eqn:three_pop}
\end{align}
where $\sum_{i=1}^n\phi_i=1$. 
The velocity distribution in Fig.\ref{fig:velocity} can be nicely fit with Eq.\ref{eqn:three_pop} and decomposed into three subpopulations ($n$=3)
with $\phi_1=0.74$, $\phi_2=0.10$, $\phi_3=0.16$, $\overline{V}_1=0.85$ $\mu m/s$, $\overline{V}_2=0.51$ $\mu m/s$, $\overline{V}_3=0.25$ $\mu m/s$, $\overline{D}_1=0.0044$ $\mu m^2/s$, $\overline{D}_2=0.0025$ $\mu m^2/s$, $\overline{D}_3=0.0043$ $\mu m^2/s$, $\overline{L}=0.75$ $\mu m$. 
This suggests that there are kinesin subpolulations with fast, intermediate, slow mean velocities.  
The peak around $\overline{V}_1$ is the most dominant, representing $\sim$ 74 \% of the population. 
As an alternative analysis method used in qunatifying the anomalous diffusion from single particle tracking \cite{Wang2009PNAS,veigel2011NRMCB}, the displacement distribution calculated over the ensemble of time traces can also reveal the three subpopulations in the ensemble (see SI and Fig.~S4).

Although the assumption that all motors move with the same mean velocity $\overline{V}$ at $t\rightarrow \infty$ is implicitly made in many studies \cite{Svoboda94PNAS,Schnitzer97Nature,Visscher99Nature,Block03PNAS}, 
variability among the motors obviously exists in the raw data. 
In the presence of motor heterogeneity, the ensemble averaged quantities, especially diffusion constant ($D$) and randomness parameter ($r$), 
are no longer well-defined quantities. 
Here the randomness $r$ is a measure of temporal regularity of motor step \cite{Svoboda94PNAS,Fisher01PNAS} (see SI):  
\begin{align}
r=\frac{\langle\tau^2\rangle-\langle\tau\rangle^2}{\langle\tau\rangle^2}=\lim_{t\rightarrow\infty}\frac{\langle x^2(t)\rangle - \langle x(t)\rangle^2}{d\langle x(t)\rangle}
\label{eqn:randomness}
\end{align} 
where $\langle\ldots\rangle$ denotes an average over the ensemble 
and $d$ is the average step size (for kinesin, $d\approx 8$ nm). 
From the second expression of Eq.\ref{eqn:randomness}, 
it is usually argued that $\langle x^2(t)\rangle -\langle x(t)\rangle^2\sim 2Dt$ and $d\langle x(t)\rangle \sim dVt$ as $t\rightarrow \infty$. 
Hence, as $t\rightarrow\infty$, the randomness, $r=2D/dV$, is ideally a time-independent, dimensionless measure of the dynamical fluctuations of the motor, or the dispersion of the motors ($x(t)$) along the track \cite{Svoboda94PNAS,Fisher01PNAS,Kolomeisky07ARPC}. 
The variation of $r$, which can be linked to the number of rate limiting steps, 
has been examined as a function of load and ATP concentration \cite{Svoboda94PNAS,Schnitzer97Nature,Visscher99Nature,Block03PNAS,Fisher01PNAS}, 
and it was concluded that kinesin data are best explained using a kinetic model with 4 states \cite{Fisher99PNAS}.
However, for these analyses based on the randomness parameter to be valid, the condition of i.i.d. should be obeyed. 

To further clarify, 
consider the position of a motor protein $\alpha$ that evolves through the following stochastic equation $\partial_t x_{\alpha}(t)= V_{\alpha}+\eta_{\alpha}(t)$,
where $V_{\alpha}$ is the mean velocity, $\langle\eta_{\alpha}(t)\rangle=0$ and 
$\langle\eta_{\alpha}(t)\eta_{\alpha}(t')\rangle=2D_{\alpha}\delta(t-t')$.
The mean square displacement (MSD) at time $t$ calculated by averaging over the motor population is (see SI): 
\begin{align}
\langle x^2(t)\rangle-\langle x(t)\rangle^2=\langle(\delta V)^2\rangle t^2+2\overline{D}t.  
\label{eqn:MSD}
\end{align}
where 
$\langle x(t)\rangle\equiv N^{-1}\sum_{\alpha=1}^Nx_{\alpha}(t)=N^{-1}\sum_{\alpha=1}^NV_{\alpha}t\equiv\overline{V}t$, 
$\langle (\delta V)^2\rangle =N^{-1}\sum_{\alpha=1}^N(V_{\alpha}-\overline{V})^2$ and $\overline{D}$ is the diffusion constant averaged over distinct ensembles.
Unless 
$\langle (\delta V)^2\rangle$ vanishes, the quadratic contribution of time becomes predominant at $t>t^*=2\overline{D}/\langle(\delta V)^2\rangle$. 
Thus, in the presence of motor heterogeneity, the non-vanishing term $\langle(\delta V)^2\rangle t^2$ renders the randomness parameter, $r$, no longer time-independent. 
Note that the Ref. \cite{Schnitzer97Nature} demonstrates the linearity of variance of position only in the short time limit.

For an ergodic system, 
 one can calculate the MSD by taking either an ensemble average over the time traces or a moving time average over a sufficiently long single time trace. 
However, 
when a system of interest has a subsample-to-subsample variation like in glassy materials, and the system is persistently heterogeneous on the physically meaningful time scale (not being able to interconvert between different components), the dynamics of the system is effectively nonergodic \cite{Palmer82AP,Lubelski08PRL}. 
A systematic procedure to analyze data with (weakly) broken ergodicity is to identify components (or domains) of the system in its configurational space as proposed by R. G. Palmer \cite{Palmer82AP, Hyeon2012NatureChem}. 

\begin{figure}[ht]
\includegraphics[width=3.2in]{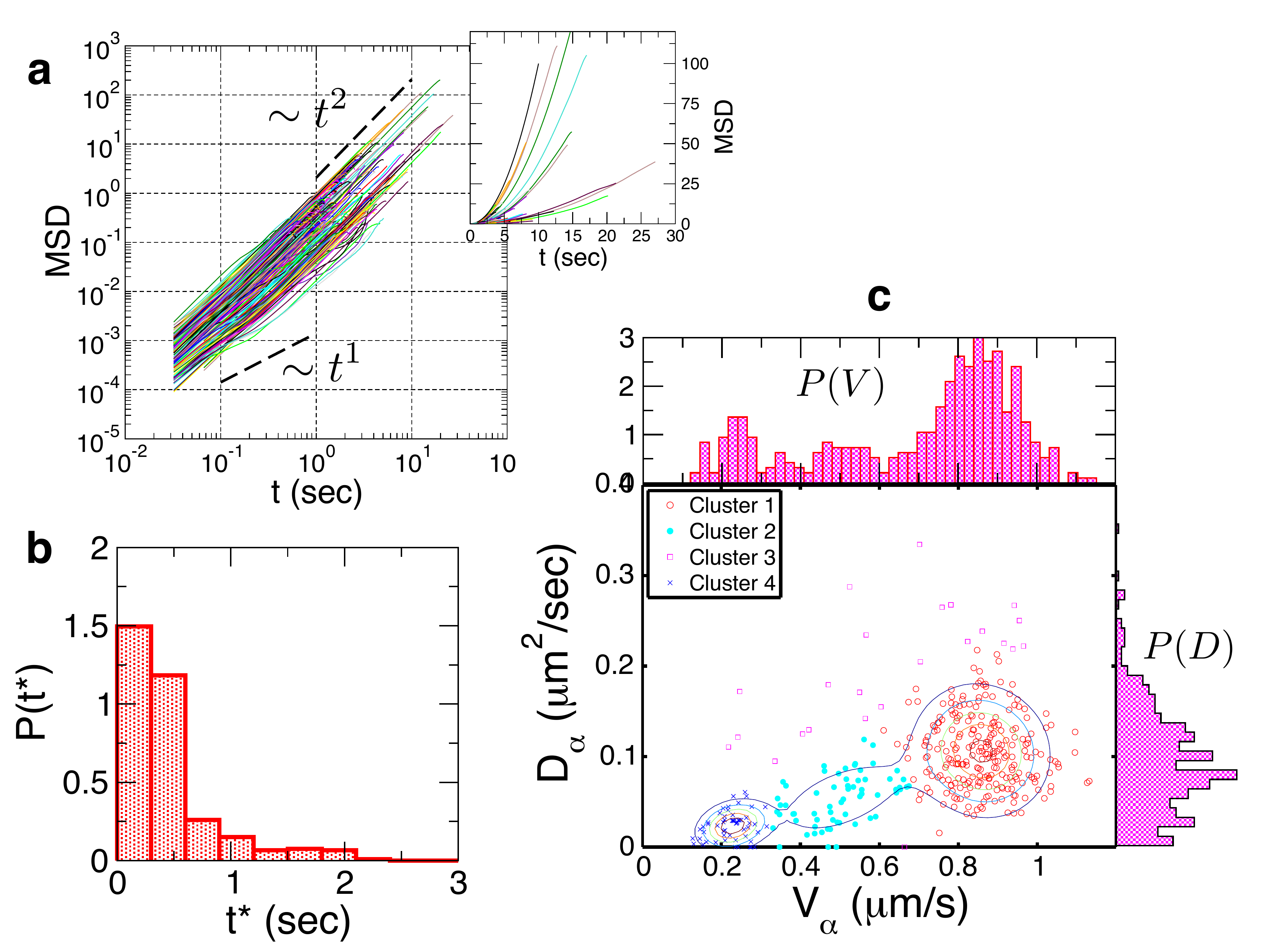}
\caption{(a) Time averaged MSD of QD time traces using Eq.\ref{eqn:MSD}. 
$\mathrm{MSD}\sim t$ for $t<2D_{\alpha}/V_{\alpha}^2$ and  $\mathrm{MSD}\sim t^2$ for $t>2D_{\alpha}/V_{\alpha}^2$. 
(b) Distribution of $t^*_{\alpha}=2D_{\alpha}/V_{\alpha}^2$. 
(c) Scatter plot of ($V_{\alpha}$,$D_{\alpha}$) and the corresponding histogram $P(V)$, $P(D)$. 
The data, partitioned into 4 clusters, are depicted in different colors. 
See Fig.S5 for the analysis of the bead assay data. 
\label{V_D_analysis.fig}}
\end{figure}

For a motor system with heterogeneity, whose ensemble properties cannot be represented with a single $V$ or $D$, one can calculate the time averaged MSD \cite{Qian91BJ,Lubelski08PRL} using
\begin{align}
\langle(\delta x_{\alpha}(t))^2\rangle_T&\equiv \frac{1}{T-t}\int^{T-t}_0\left(x_{\alpha}(t'+t)-x_{\alpha}(t')\right)^2dt'\nonumber\\
&=V_{\alpha}^2t^2+2D_{\alpha}t
\label{eqn:MSD}
\end{align}
with $T\gg t$ for each molecule $\alpha$, and quantify both the directed ($V_{\alpha}$) and the diffusive ($D_{\alpha}$) components. 
The log-log plot shows MSD$\sim t$ for $t<t_{\alpha}^*$ and MSD$\sim t^2$ for $t>t_{\alpha}^*$ (Fig.\ref{V_D_analysis.fig}a); a crossover from diffusive to directed dynamics is observed at $t_{\alpha}^*=2D_{\alpha}/V_{\alpha}^2$ (Fig.\ref{V_D_analysis.fig}b).  
In addition to the $P(V)$ in Fig.\ref{fig:velocity}c, 
the scatter plot ($V_{\alpha}$,$D_{\alpha}$) (Fig.\ref{V_D_analysis.fig}c) clearly visualizes the presence of heterogeneity with multiple subpopulations. 
Using a multivariate Gaussian mixture model \cite{press2007numerical} we partitioned ($V_{\alpha}$,$D_{\alpha}$) into 4 clusters. 
When the outlier subpopulation (cluster 3) is excluded, the remaining three subpopulations show a weak correlation between $V_{\alpha}$ and $D_{\alpha}$. 
The scatter plot of ($V_{\alpha}$, $D_{\alpha}$) suggests that kinesins with higher mean velocity tend to have more variation in stepping, reminiscent of the recent finding that the heat released during catalytic turnover enhances the enzyme diffusion \cite{Riedel2014Nature}.

To summarize, our study provides simple criteria to judge the homogeneity of motor time traces:  
(i) Time traces with large run length should have a converging velocity.  
$P(V)$ ought to be (ii) uni-modal, (iii) symmetric with respect to its mean.  and (iv) described by $P_{\text{homo}}(V)$ (Eq.\ref{eqn:PV}) (or should satisfy FWHM $\approx 3.1(\overline{V}\overline{D}/\overline{L})^{1/2}$).  
If a distribution satisfies the criteria (ii) and (iii), yet heterogeneity is still suspected then one can check the criterion (iv) (Statistical assessment of a possible difference between the two distributions, $P(V)$ and $P_{\text{homo}}(V)$ can be made, for example, using the Kolmogorov-Simirnov test). 
Violation of the above criteria most likely indicates the presence of motor heterogeneity. 
Once heterogeneity is identified, we propose to decompose the heterogeneous ensemble of kinesin motors into a finite number of homogeneous sub-ensembles. 

Care should be taken in developing theories for heterogenous samples, as has been pointed out in the problem of phase separations of practical polymer samples with polydispersity \cite{Koningsveld70FaradaySoc,deGennesbook}.  
In biology, although importance of heterogeneity is increasingly appreciated at the cellular level \cite{altschuler2010cell,junker2014Cell}, 
it is surprising to observe such characteristics at the level of single biomolecules. 
The structural origin of functional heterogeneity or multiple native states \cite{Liu2013Nature,Solomatin10Nature} is not clear. 
Yet, all the data from different control experiments discussed here unambiguously indicate that kinesins can have multiple functional states with distinct capability of processing ATP, 
which gives rise to fast, intermediate, and slow groups of motors.  
It is plausible that while the conformational cycle of kinesin is ``driven" by chemical potential from ATP hydrolysis, functional states of motile kinesins are dynamically pinned or separated by large free energy barrier over which no thermodynamic path can easily connect one state with another.  
Although it is not easy to test this hypothesis for kinesins due to the short processivity, recent studies of Holliday junctions \cite{Hyeon2012NatureChem} and RecBCD \cite{Liu2013Nature} have shown that subensemble-to-subensemble interconversion can be induced by depleting cofactors (Mg$^{2+}$ and ATP, respectively) for a finite amount of time. 

Regardless of the cause of the heterogeneity, it is important to appropriately characterise it, because one goal of single molecule studies is to measure properties to allow prediction/calculation of ensemble function. Ensembles of heterogeneous molecules are likely to function quite differently from ensembles of homogeneous ones. 
For instance, heterogeneous motors may interfere more with each other, and in general systems combining multiple motors with different velocities, unexpected ensemble properties can emerge \cite{klumpp2005PNAS,gross2007CurrBiol,berger2012PRL,li2013BJ}.

\begin{acknowledgments}
CH thanks Anatoly Kolomeisky for useful discussion and the KITP at the University of California, Santa Barbara (Grant No. NSF PHY11-25915), for support during the preparation of the manuscript.
SPG was supported by NIH RO1 GM070676 
\end{acknowledgments}

\bibliographystyle{apsrev}
\bibliography{mybib1}

\clearpage 

\section{Supplementary Information}
\makeatletter 
\renewcommand{\thefigure}{S\@arabic\c@figure}
\renewcommand{\theequation}{S\@arabic\c@equation}
\makeatother 
\setcounter{figure}{0}  
\setcounter{equation}{0}
{\bf Experimental Methods.}

{\it Protein:} 
To ensure that the population of kinesins are chemically ``identical", we expressed the functional, truncated kinesin (K560) \cite{woehlke1997Cell} in terrific broth and purified it as described in Ref.\cite{woehlke1997Cell,xu2012NatureComm}. 
Note that the lack of light chains (tails) prevents potential interactions between light and heavy chains that may alter function. 
Further, we minimized the chance of phosphorylation, another potential source of chemical heterogenity, by expressing the kinesins in E. coli. 
Furthermore, to select only the functional kinesins, the stock protein was selectively purified via MT binding and release in the presence of AMP-PNP (the nonhydrolizable ATP analogue). 
\\

{\it Quantum Dot (QD) Motility Experiment:} Truncated kinesin-1 was specifically recruited to quantum dots via its genetically encoded C-terminal His-tag \cite{lu2009Traffic}.  To achieve the specific linking of kinesin to the cargo, streptavidin quantum dots (QD-655-streptavidin conjugate, Life Technologies) were labeled with biotin conjugated penta-his antibody (Qiagen) in molar ratio of 1:1::AB:QD. The incubation temperature of antibody with QDs was 4C and lasted for 1 hour. The AB labeled QD surface was blocked with 4mg/ml casein (sigma Aldrich, C8654-500G) in the motility buffer (80 mM Pipes pH 6.9, 50 mM CH3CO2K, 4 mM MgSO$_4$, 1 mM DTT, 1 mM EGTA, 10 $\mu$M taxol, 4mg/ml casein) for 1h at 4C to reduce nonspecific binding. With this blocking procedure K560 binding to casein coated streptavidin QDs alone (without antibody) was negligible (reduced by 50 times) compared to the same for AB coated QDs. 
K560 with His tag on its truncated tail was incubated with anti-His antibody tagged QDs in the molar ratio of 1:40 at RT for 12 minutes. Before using the mixture to test motility it was supplemented with 1 mM ATP and oxygen-scavenging system. The high ratio of K560:QD (=1:40) was chosen to ensure the likelihood of recruiting a single kinesin to the QDs \cite{xu2012NatureComm}. A sample chamber assembled with taxol stabilized microtubules made from bovine brain tubulin (cytoskeleton) was used for motility experiments. 
Consistent with the hypothesis that individual QDs were at most attached to a single kinesin, the distribution of run-lengths was well fit by a single decaying exponential distribution, with a mean travel of 0.78 microns (Fig. S2). This travel distance is what would be expected for single-motor travel. 
\\

{\it TIRF Imaging and Tracking:} The QDs were imaged via custom built objective based total internal reflection fluorescence microscope (Nikon 1.49NA, 100X, and 488 nm laser Ti:Sapphire, Coherent). The time lapse movies were recorded at 31fps using EMCCD camera (Photmetrics QuantEM 512SC).  Tracking analysis was carried out using a custom matlab program (Gross Lab) that identified QD positions in the images via 2D Gaussian fitting of their intensity profile.
\\

{\it Levitated microtubule assays:} 
To eliminate concerns about surface interactions, we developed Ôelevated microtubuleÕ assay. In this assay 500nm carboxyl terminated polystyrene beads were incubated with enzymatically dead kinesin (a mutant, E237A in hKIF5A) at RT for 5 minutes. The single E237A mutation in the enzymeÕs switch 2 region precludes ATP hydrolysis, rendering the motor immobile even in the presence of saturating ATP concentrations thus acting like a MT anchor. After recruitment the beads surface was then blocked with casein, and the beads were spun down at 5000g and re-suspended in casein buffer, to eliminate free kinesin in from buffer. These beads were then flown into a flow cell, and allowed to stick to the pre-cleaned polylysine coated coverslips for 6 minutes. At the end of 6 minutes the new buffer with high casein levels (5.5 mg/ml) was flowed in, to reduce nonspecific binding of MTs and QDs to coverslip surface. Casein coating made the surface minimally reactive, so that when taxol-stabilized microtubules were flown into the coverslip, they would not stick to the coverslip, and could easily be washed out if the beads with mutant kinesin were not present. However, in the presence of mutant kinesin coated beads, taxol-stabilized MTs ended up being ÔcapturedÕ by the mutant kinesins on the bead, so that the MTs were suspended between beads, but above the surface. As this was a stochastic process, some MTs ended up close to the surface, and some further away. 

\begin{figure*}[h]
\includegraphics[width=5.4in]{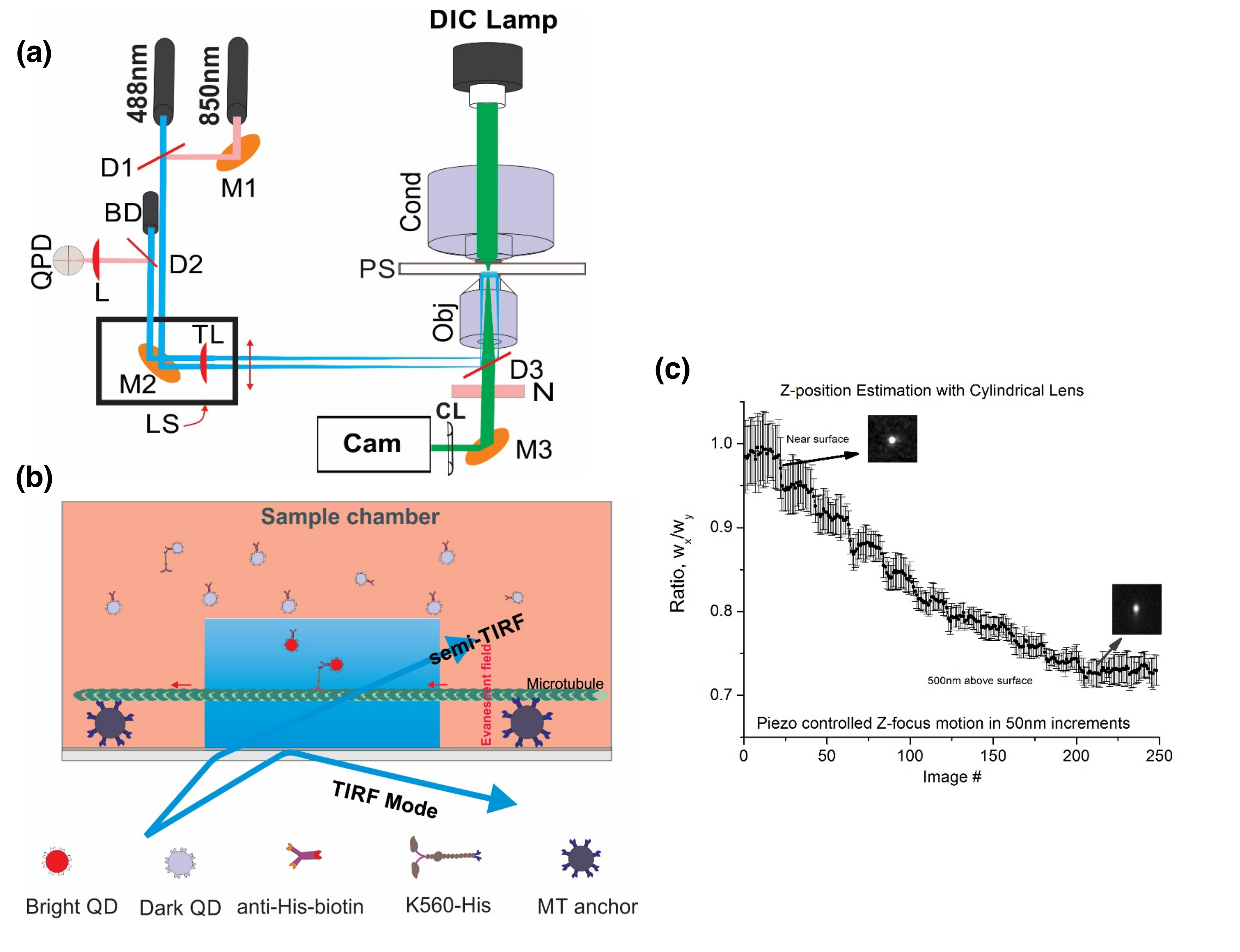}
\caption{
Geometry and calibration of off-surface measurements.  
(a) Diagram of experimental apparatus, showing location of cylindrical lens (CL) in front of camera, to induce distortion when the QD is out of focus. 
(b) Diagram of experimental geometry. Half-micron beads coated with mutant  kinesin (which rigor-binds to MTs) are attached to the coverslip, and microtubules are subsequently flowed in, and stick to the beads, ending up suspended between beads  above the surface.  
(c) Quantification of asymmetry in QD image, as a function of the QDs distance from the plane of focus. The extent of asymmetry was used to detect QDs moving on MTs either close to or far-from the surface. 
The error bars are SEM, estimated by tracking the position and intensity profiles of 20 QDs in the time lapse images recorded during piezo $Z$ motion.
\label{laser.fig}}
\end{figure*}

Once the microtubules were in place, we now flowed in QD-labeled kinesin motors (as above) and recorded their motion via wide-field semi-TIRF imaging, as shown in Fig. \ref{laser.fig}b.  
Importantly, since the MTs were at different heights above the surface, to measure the height (distance from surface) of a particular QD, we used the elipticity in the point spread function of the QDs (Fig.\ref{laser.fig}c). 
In brief, by putting in a cylindrical lens (f=1000mm, at ~80mm from CCD chip) in front of the camera, we could induce astigmatism similar to 3 dimensional STORM imaging \cite{Huang2008NatureMethods}; by quantifying the amount of distortion for a given QD, we could determine its height. To calibrate this system, the detector was focused on the coverslip surface, and then the surface was moved in known increments in the $Z$ direction (via a piezo-controlled stage), and we quantified how the images of surface-fixed QDs changed with $Z$ displacement (Fig. \ref{laser.fig}c).  
For our studies the elipticity of the ``off-surface" QDs reflects an average QD height of 250 nm. 
\\

\begin{figure}[h]
\includegraphics[width=2.8in]{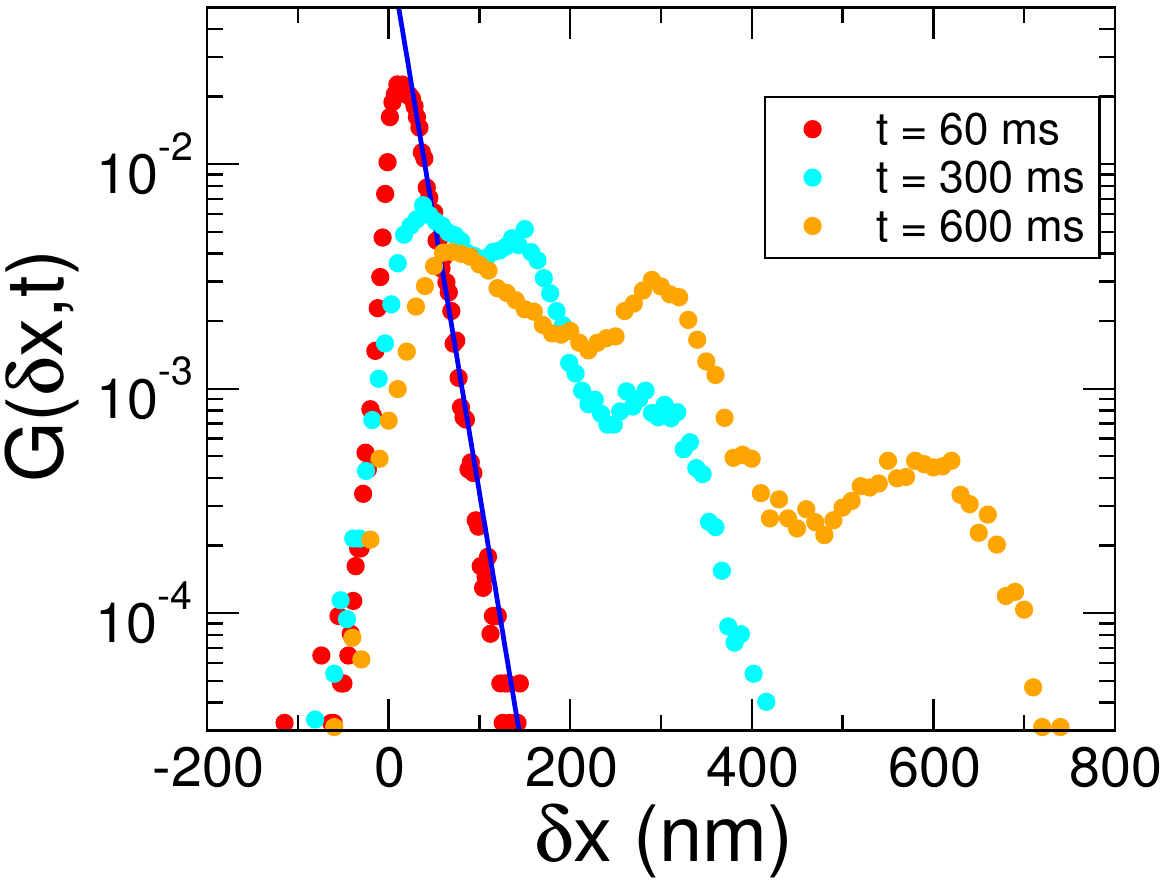}
\caption{Displacement distribution $G(\delta x,t)$ calculated over the ensemble of kinesin time traces for three different time interval $t$. 
The tail region ($\delta x\gg 1$) of $G(\delta x, t=60\text{ ms})$ is described with an exponential function. 
\label{displacement_distribution.fig}}
\end{figure}

{\bf Analysis of kinesin time traces using displacement distribution.}
The displacement probability distribution, a quantity often calculated for the purpose of revealing the anomalous diffusion in single particle tracking, could be used as an alternative method for analyzing our \emph{in vitro} kinesin data. The relative increment of displacement measured with a uniform time interval of 60 milisecond calculated for the ensemble of our QD-labeled kinesins on surface-immobilized MTs shows non-gaussian distribution, fitted to an exponential for large displacements (see Fig.\ref{displacement_distribution.fig}). 
The exponential tail is similar to the data analyzed for the colloidal beads diffusing on lipid tubes or particles diffusing in F-actin network in Ref.\cite{Wang2009PNAS}. What is even more interesting is the displacement distributions calculated at larger time intervals ($t = 300$ ms and 600 ms. See Fig.\ref{displacement_distribution.fig}), which apparently capture the presence of the three distinct components in the ensemble of the kinesin time traces. Since the displacement at a large time interval effectively becomes equivalent to the global velocity measured for individual motors, the three peaks observed in $G(\delta x,t)$ at large $t$ indicate that there are three distinct subpopulations in the ensemble of kinesin motors.  
\\

{\bf Meaning of the randomness paramter, $r$.}
From the definition of the randomness parameter $r$
\begin{equation}
r=\frac{\langle\tau^2\rangle-\langle\tau\rangle^2}{\langle\tau\rangle^2}
\end{equation}
and $\langle\tau^n\rangle=\int^{\infty}_0dtt^n\psi(t)$, 
an exponential poisson process $\psi(t)=\tau^{-1}e^{-t/\tau}$ leads to $r=1$; and $\psi(t)=\tau_n^{-1}t^ne^{-t/\tau_n}/n!$ to $r=1/(n+1)$.
If $r\rightarrow 0$, there is little variation in each dwell time; thus the motor step is clock-like. By contrast, if $r\rightarrow 1$, the duration of individual steps is highly irregular. 
\\

{\bf Derivation of Eq.5.} 
Consider a motor protein $\alpha$ that evolves through the stochastic equation $\partial_t x_{\alpha}(t)= V_{\alpha}+\eta_{\alpha}(t)$, where $V_{\alpha}$ is the mean velocity, $\eta_{\alpha}(t)$ is the Gaussin random noise with zero mean. 
From the formal solution of $x_{\alpha}(t)=V_{\alpha}t+\int^t_0d\tau\eta_{\alpha}(\tau)$ with $x_{\alpha}(0)=0$,  
the ensemble averaged displacement is $\langle x(t)\rangle=\overline{V}t$ where $\langle x(t)\rangle\equiv\frac{1}{N}\sum_{\alpha=1}^Nx_{\alpha}(t)$ is the ensemble average displacement of motors at time $t$, and $\overline{V}=\frac{1}{N}\sum_{\alpha=1}^NV_{\alpha}$. 
Then, the mean square displacement (MSD) for the motor ensemble is: 
\begin{widetext}
\begin{align}
\langle x^2(t)\rangle-\langle x(t)\rangle^2 &\equiv \frac{1}{N}\sum_{\alpha=1}^N(x_{\alpha}(t)-\langle x(t)\rangle)^2\nonumber\\
&=\frac{1}{N}\sum_{\alpha=1}^N(V_{\alpha}-\overline{V})^2t^2+\frac{1}{N}\sum_{\alpha=1}^N\int^t_0d\tau_1\int^t_0d\tau_2\eta_{\alpha}(\tau_1)\eta_{\alpha}(\tau_2)\nonumber\\
&=\langle(\delta V)^2\rangle t^2+2\overline{D}t. 
\label{eqn:MSD_derivation}
\end{align}
where we define the variance of the mean velocities as $\langle (\delta V)^2\rangle \equiv\frac{1}{N}\sum_{\alpha=1}^N(V_{\alpha}-\overline{V})^2$. 
For the noise related term from the second to the third line of Eq.\ref{eqn:MSD_derivation}, we use the property of Gaussian random noise, whose correlation satisfies $\frac{1}{N_p}\sum_{i=1}^{N_p}\eta_i(\tau_1)\eta_i(\tau_2)=2D_p\delta(\tau_1-\tau_2)$, where the subscript $p$ denotes the index of a subensemble ($p=1,\ldots, n_c$); hence, 
\begin{align}
\frac{1}{N}\sum_{\alpha=1}^N\int^t_0d\tau_1\int_0^td\tau_2\eta_{\alpha}(\tau_1)\eta_{\alpha}(\tau_2)
&=\frac{1}{n_c}\sum_{p=1}^{n_c}\int^t_0d\tau_1\int_0^td\tau_2\frac{1}{N_p}\sum_{i=1}^{N_p}\eta_i(\tau_1)\eta_i(\tau_2)\nonumber\\
&=\frac{1}{n_c}\sum_{p=1}^{n_c}\int^t_0d\tau_1\int_0^td\tau_22D_p\delta(\tau_1-\tau_2)\nonumber\\
&=2\overline{D}t
\end{align}
where $\sum_{p=1}^{n_c}\sum_{i=1}^{N_p}1=N$, and $\overline{D}=\frac{1}{n_c}\sum_{p=1}^{n_c}D_p$ is the diffusion constant averaged over the distinct subensembles. 
\end{widetext}

\begin{figure*}[h]
\includegraphics[width=6.5in]{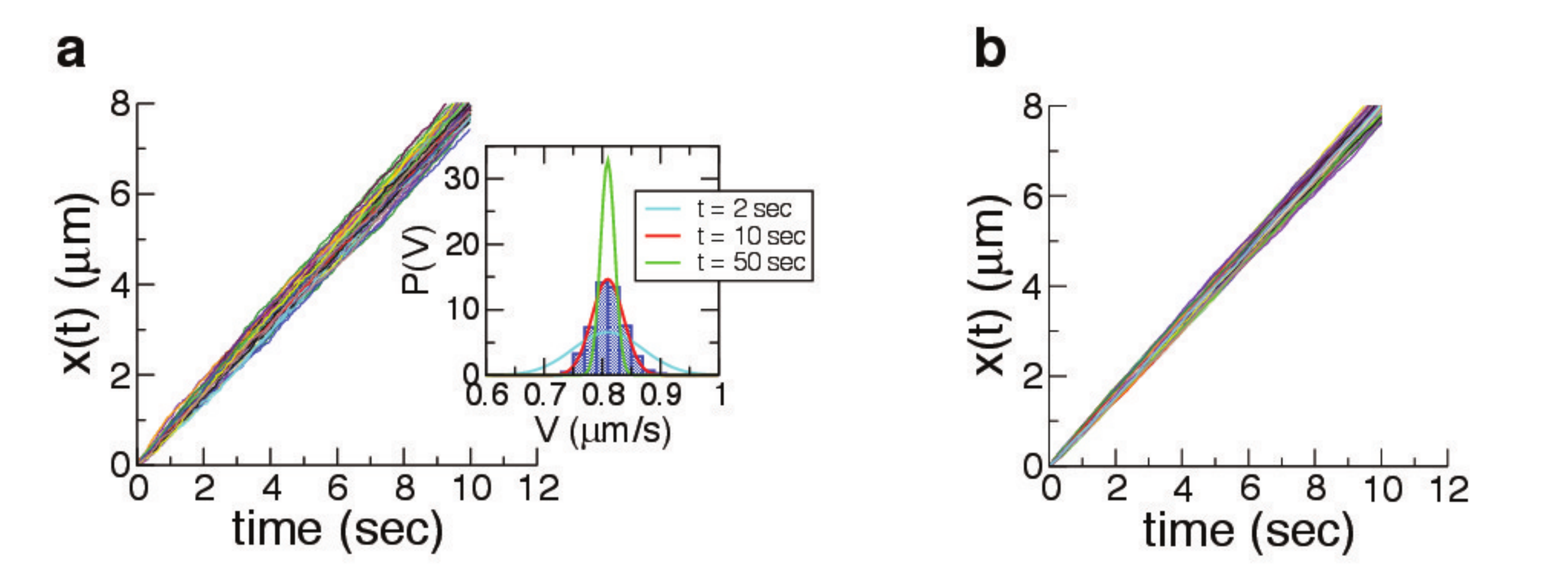}
\caption{Computer-generated time traces ($N=1000$) by using stepping time distributions: (a) $\psi(t)=\tau^{-1}e^{-t/\tau}$ and (b) $\psi(t)=(\tau_5)^{-1}t^5e^{-t/\tau_5}/5!$ where $\tau_5\equiv\tau/5$ with $\tau=10$ ms and step size of $d=8$ nm. 
The histogram of mean velocity (inset of (a)) is fitted to Eq.1 in the main text with $\overline{V}=0.8$ $\mu m/s$ and $\overline{D}=0.0032$ $\mu m^2/s$ at $t=10$ sec. $P(V)$s at $t=2$ and 50 sec are also plotted.  
\label{Gaussian.fig}}
\end{figure*}

\begin{figure}[h]
\includegraphics[width=2.8in]{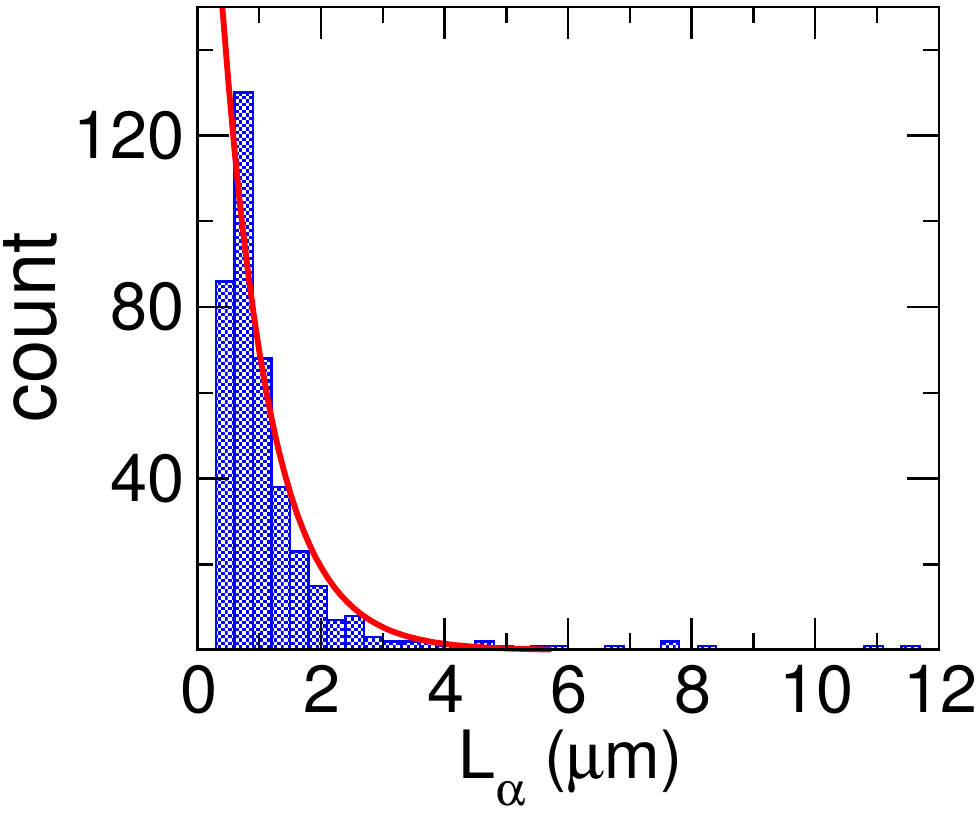}
\caption{Histogram (unnormalized) of travel distances from QD assays. 
The decay length from the single-exponential fit was $\overline{L}=0.78$ $\mu m$.
\label{histogram_unnormalized.fig}}
\end{figure}

\begin{figure}[h]
\includegraphics[width=3.4in]{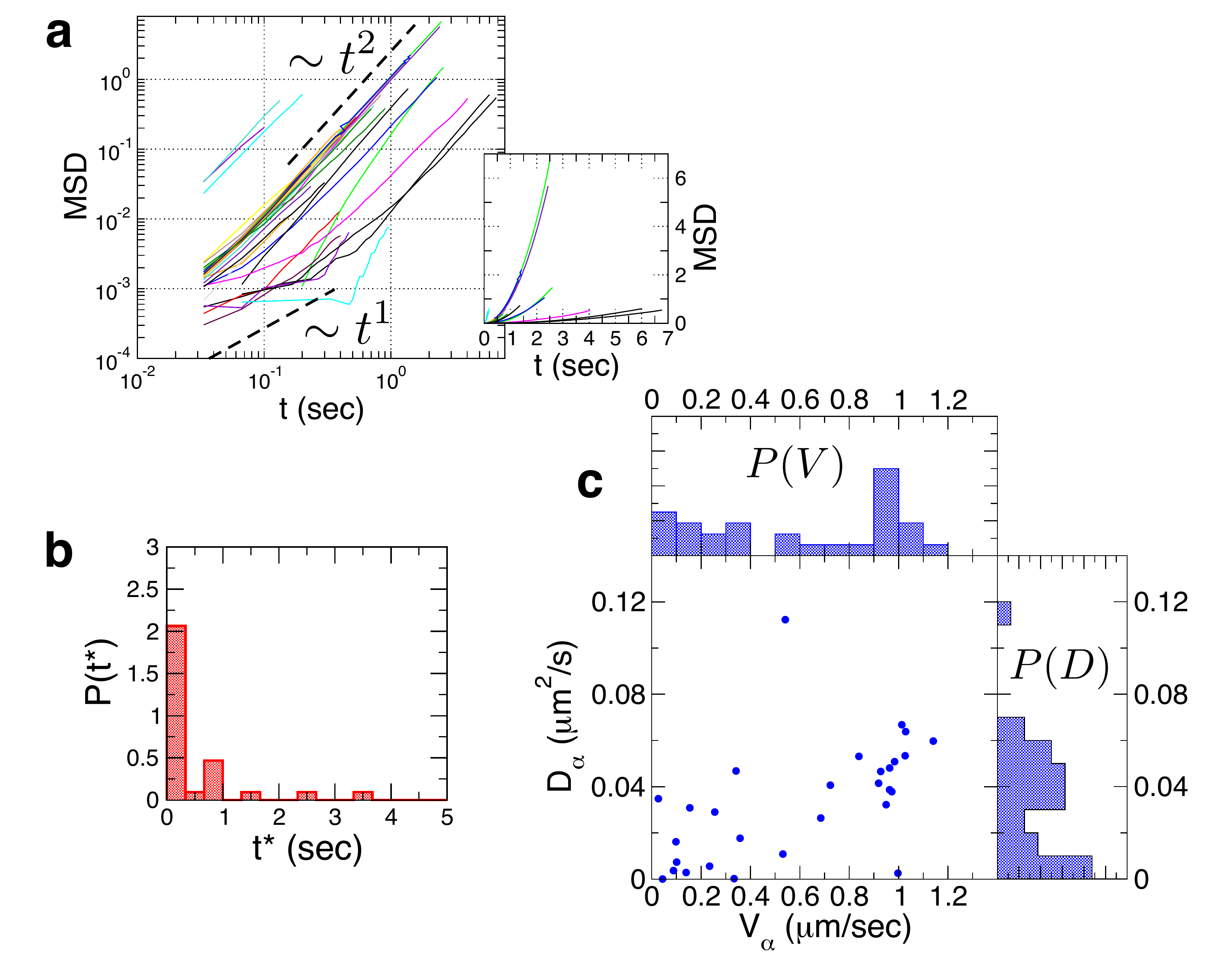}
\caption{MSD analysis for bead assays with specific attachment. (a) MSD of time traces from temporal moving average defined in Eq.6. 
MSD versus time in logarithmic scale. $\mathrm{MSD}\sim t$ for $t<2D_{\alpha}/V_{\alpha}^2$ and  $\mathrm{MSD}\sim t^2$ for $t>2D_{\alpha}/V_{\alpha}^2$. 
(b) Distribution of the cross-over time $t^*_{\alpha}=2D_{\alpha}/V_{\alpha}^2$. 
(c) Scatter plot of ($V_{\alpha}$,$D_{\alpha}$) and the corresponding histogram $P(V)$, $P(D)$. 
\label{V_D_analysis_bead_specific.fig}}
\end{figure}

\end{document}